\documentclass[showpacs,showkeys,superscriptaddress,aps]{revtex4}
\begin{document}
\title{The nature of manifolds of periodic points for higher dimensional integrable maps II}
\author{Satoru SAITO}
\email[email : ]{saito_ru@nifty.com}
\affiliation{Hakusan 4-19-10, Midori-ku, Yokohama 226-0006 Japan}
\author{Noriko SAITOH}
\email[email : ]{nsaitoh@ynu.ac.jp}
\affiliation{Department of Applied Mathematics, Yokohama National University\\
Hodogaya-ku, Yokohama, 240-8501 Japan}
\begin{abstract}
We study periodicity conditions of a rational map on $\bm{C}^d$ with $p$ invariants and show that a set of isolated periodic points and an algebraic variety of finife dimension do not exist in one map simultaneously if $p\ge d/2$. We also discuss in detail how the transition takes place between them.
\end{abstract}
\maketitle
\section{Introduction}

By studying transitions of higher dimensional nonintegrable maps to integrable ones, we showed in our previous paper\cite{SSI} that the periodicity conditions, which are `uncorrelated' in generic systems, become `fully correlated' in the integrable maps and one gets full manifolds, instead of isolated points, characterized by invariants of the map. The purpose of this paper is to establish a criterion which distinguishes the fully correlated and uncorrelated periodicity conditions. We will also discuss in detail how the transition takes place between them.

We consider a rational map on $\bm{C}^d$
\begin{equation}
\bm{x}=(x_1,x_2,...,x_d)\quad \rightarrow\quad
\bm{X}=(X_1,X_2,...,X_d)=\bm{X}^{(1)}.
\label{x->X}
\end{equation}
We are interested in the behaviour of the sequence: $
\bm{x}\rightarrow \bm{X}^{(1)}\rightarrow\bm{X}^{(2)}\rightarrow\cdots$. In the study of the sequence of the map, the periodic points of the map decide the most important nature of the sequence. To see their role let us begin our discussion by introducing some words. The precise meaning of them will be explained in \S 3.

Let us assume that the map has $p$ invariants. The periodicity conditions are called `uncorrelated' if they determine all positions of the periodic points dependent on the values of the invariants. They are called `fully correlated' if they do not determine the position of the periodic points but impose relations on the values of the invariants. An `invariant variety of periodic points' is a variety determined by the invariants of the map alone, such that every point on the variety can be an initial point of the iteration, which stays on the variety before it returns to the same point. We have shown in \cite{SSI} the following propositions:

\begin{itemize}
\item {\it A set of periodic points form a set of isolated points if the periodicity conditions are uncorrelated, and an invariant variety of dimension $p$ if they are fully correlated.}
\item {\it An invariant variety of periodic points and a set of isolated periodic points do not exist in one map simultaneously.}
\end{itemize}

We would like to prove the following theorem in this paper, from which the above second proposition immediately follows.\\

\noindent
{\bf Theorem:}

\begin{itemize}
\item
{\it When $p<d/2$, there is no set of fully correlated periodicity conditions.}
\item
{\it When $p\ge d/2$, sets of correlated and uncorrelated periodicity conditions do not exist in one map simultaneously.}
\end{itemize}

We don't know exactly the correspondence between the existence of an invariant variety of periodic points and integrability of the map. Nevertheless we have presented in \cite{SSI} many examples of invariant variety found in some known integrable rational maps. On the other hand, the isolated periodic points, which characterize non-integrable maps, are derived by solving uncorrelated periodicity conditions. Therefore it will be worthwhile to study the transition of periodicity conditions from uncorrelated regime to fully correlated one. This will be our second topic in this paper. We would like to discuss in detail a simple example and see how the transition takes place when a parameter introduced to the map approaches from one regime to the other. In particular we focuss our attention to the behaviour of the Julia set, the closure of the set of repulsive periodic points of a map.\\

This paper will be organized as follows. We study in \S 2 a general scheme of higher dimensional maps with $p$ invariants. The scheme enables us to investigate integrable and nonintegrable maps on an equal footing. The proof of the theorem will be given in \S 3. Based on the results of the previous sections we develop in \S 4 a method generating invariant varieties of periodic points for some simple cases. We study in \S 5 how the transition takes place between the uncorrelated and fully correlated periodicity conditions.

\section{Dimensional reduction by invariants}

The integrability of a dynamical system with continuous time evolution is guaranteed by the existence of sufficient number of invariants of motion due to the Liouville-Arnold theorem. There is, however, no corresponding theorem in the case of iteration of maps. In this section we study the iteration of a map (\ref{x->X}) on $\bm{C}^d$ with $p$ invariants. 

To begin with we study a general scheme of a map with $p$ invariants $H_i(\bm{x}),\ i=1,2,...,p$. If the functions $f_j(\bm{x}),\ \ j=1,2,...,d$ satisfy
\begin{equation}
H_i(f_1(\bm{x}),\cdots, f_d(\bm{x}))=H_i(x_1,x_2,\cdots,x_d),,\qquad i=1,2,...,p
\end{equation}
the $d$ dimensional map
\begin{equation}
(x_1,x_2,\cdots,x_d)\quad \rightarrow\quad (X_1,X_2,\cdots,X_d)=\Big(f_1(\bm{x}),f_2(\bm{x}),\cdots,f_d(\bm{x})\Big)
\label{map}
\end{equation}
restricts to any common level set of $H_1,\cdots,H_{p}$. More generally, the invariants $H_i$ may be specified implicitly, for example by
\begin{equation}
\varphi_i\Big(f_1(\bm{x}),f_2(\bm{x}),\cdots, f_d(\bm{x}),H_i\Big)=0,\qquad i=1,2,...,p.
\label{varphi_j(f_1,f_2,cdots, f_d,h_j)=0}
\end{equation}
In general the equations (\ref{varphi_j(f_1,f_2,cdots, f_d,h_j)=0}) fix a level set $V(h)$,
\begin{equation}
V(h)=\Big\{\bm{x}\Big|\ H_i(\bm{x})=h_i,\ i=1,2,...,p\Big\},
\label{V(h)}
\end{equation}
with $h_1,h_2,...,h_p$ the values of the invariants determined by the initial data of the map. $V(h)$ is a $d-p$ dimensional variety in $\bm{C}^d$ and (\ref{map}) defines a map whose trajectory of iteration is constrained to the variety. Different choices of $f_1$ correspond to different parametrizations of the trajectory.\\

As an example, let $H(x,y)=x+y$, {\it i.e},
\begin{equation}
\varphi(x,y,H)=x+y-H.
\label{varphi(x,y,h)=x+y-h}
\end{equation}
If we fix $f_1(x,y)$ by
\begin{equation}
f_1(x,y)=xy,
\label{f_1(x,y)=xy}
\end{equation}
then, by solving $f_1+f_2=x+y$, we find $f_2(x,y)=x+y-xy$, and the map
\begin{equation}
(x,y)\quad \rightarrow\quad (X,Y)=\Big(xy,\ x+y-xy\Big)
\label{X=xy,Y=x+y-xy}
\end{equation}
restricts to any line $x+y=h$. Using this fact we can reduce the map (\ref{X=xy,Y=x+y-xy}) to one dimension,
$$
X=x(h-x),\qquad Y=h-X.
$$
We see that the 2-dim map (\ref{X=xy,Y=x+y-xy}) is equivalent to a logistic map, hence is not integrable although it has the invariant $H=x+y$.

Next, consider an example where the level set is a hyperbola:
\begin{equation}
\varphi(x,y,H)=(1-bx)y-(1-cx)H,
\label{(1-bx)y-(1-cx)h}
\end{equation}
and fix the same function (\ref{f_1(x,y)=xy}) for $f_1(x,y)$. We obtain
$$
f_2(x,y)=H{1-cf_1(x,y)\over 1-bf_1(x,y)}=y{1-bx\over 1-cx}\ {1-cf_1(x,y)\over 1-bf_1(x,y)}
$$
and hence the map
\begin{equation}
(x,y)\quad \rightarrow\quad (X,Y)=\left(xy,\ y{(1-bx)(1-cxy)\over (1-cx)(1-bxy)}\right).
\label{2d map}
\end{equation}
It is straightforward to verify that
\begin{equation}
H(x,y)=y{1-bx\over 1-cx}
\label{H=y(1-bx)/(1-cx)}
\end{equation}
is an invariant.

If we denote by $h$ the constant of the map, the reduction to one dimension yields 
\begin{equation}
X=hx{1-cx\over 1-bx}.
\label{X=hx(1-cx)/(1-bx)}
\end{equation}
The other component $Y$ is given, once $X$ is known, by the formula $Y=h(1-cX)/(1-bX)$.

(\ref{X=hx(1-cx)/(1-bx)}) is apparently nonintegrable when $b=0$ and $c\ne 0$ since it is a logistic map. It is, however, integrable when $c=0$ or $c=b$ since it becomes a M\"obius map. Note that, if we had chosen a different $f_1$, say $x+y$ instead of $xy$, we obtained a nonintegrable map with the same invariant even at $c=0$. We study the map (\ref{2d map}) in detail in \S 4 and 5.\\

For the third example we consider the case where the level set is a biquadratic curve defined by
\begin{equation}
\varphi(x,y,H)=ay^2x^2+b(y+x)yx+c(y-x)^2+dyx+e(y+x)+f,
\label{biquadratic curve}
\end{equation}
or equivalently
\begin{equation}
\varphi(x,y,H)=\phi(x)y^2+\eta(x)y+\rho(x),
\label{phi(x)y^2+eta(x)y+rho(x)}
\end{equation}
where
\begin{equation}
\phi(x):=ax^2+bx+c,\quad \eta(x):=bx^2+(d-2c)x+e,\quad
\rho(x):=cx^2+ex+f.
\label{phi,eta,rho}
\end{equation}
We have not written $H$ explicitly but assume that it is hidden in the parameters 
$$
\bm{q}=(a,b,c,d,e,f)
$$
which characterize the curve.

If we fix $f_1(x,y)$ simply by
\begin{equation}
f_1(x,y)=y,
\label{f_1}
\end{equation}
we obtain $f_2(x,y)$ from (\ref{phi(x)y^2+eta(x)y+rho(x)}) as
\begin{equation}
f_2(x,y)={-\eta(y)\pm\sqrt{\eta^2(y)-4\phi(y)\rho(y)}\over 2\phi(y)}.
\label{f_2}
\end{equation}
This particular choice of $f_1(x,y)$ defines a one dimensional map $x\rightarrow X$ specified by the equation
\begin{equation}
S(X,x;\bm{q})=0
\label{S(X,x)=0}
\end{equation}
where
\begin{equation}
S(X,x;\bm{q}):=aX^2x^2+b(X+x)Xx+c(X-x)^2+dXx+e(X+x)+f.
\label{S(X,x)}
\end{equation}
We shall call a map of this form a `biquadratic map' in the following.

The image of the map $(X,Y)$ is determined if $H$ is given. As an example let us assume that
\begin{equation}
\bm{q}=\bm{q}'+h\bm{q}'',
\label{p=p'+p''}
\end{equation}
and try to find a map which conserves $H=h$. Substituting (\ref{p=p'+p''}) into (\ref{biquadratic curve}) or (\ref{phi(x)y^2+eta(x)y+rho(x)}) we see that the quantity which must be conserved is
\begin{equation}
H(x,y)=-\ {\phi'(x)y^2+\eta'(x)y+\rho'(x)\over \phi''(x)y^2+\eta''(x)y+\rho''(x)}.
\label{h QRT}
\end{equation}
Here $(\phi',\eta',\rho')$ and $(\phi'',\eta'',\rho'')$ are defined by changing $\bm{q}$ in $(\phi,\eta,\rho)$ to $\bm{q}'$ and $\bm{q}''$, respectively. According to our prescription of finding a map the image $(X,Y)$ of $(x,y)$ is derived by the replacement of $h$ in $(f_1,f_2)$ by (\ref{h QRT}). After some manipulation we obtain the map
\begin{equation}
(x,y)\quad\rightarrow\quad (X,Y)=\left(y,\ {(\eta'\rho''-\rho'\eta'')(y)-x(\rho'\phi''-\phi'\rho'')(y)\over
(\rho'\phi''-\phi'\rho'')(y)-x(\phi'\eta''-\eta'\phi'')(y)}\right).
\label{QRT 2d map}
\end{equation}
If we write $(Y,y,x)$ as $(x_{n+2},x_{n+1},x_{n})$, this is nothing but the symmetric version of the well known QRT equation\cite{QRT},
\begin{equation}
x_{n+2}={(\eta'\rho''-\rho'\eta'')(x_{n+1})-x_{n}(\rho'\phi''-\phi'\rho'')(x_{n+1})\over
(\rho'\phi''-\phi'\rho'')(x_{n+1})-x_{n}(\phi'\eta''-\eta'\phi'')(x_{n+1})}.
\label{QRT equation}
\end{equation}
From our construction the correspondence of this two to one map $(x_n,x_{n+1})\rightarrow x_{n+2}$ to the biquadratic map (\ref{S(X,x)=0}) is obvious. \\

Similarly we can use the general scheme to study higher dimensional maps. For example we define
\begin{equation}
\varphi_1(x,y,z,H_1)=xyz-H_1,\qquad \varphi_2(x,y,z,H_2)=(1-x)(1-y)(1-z)-H_2,
\label{r,s}
\end{equation}
and fix $f_1$ by
$$
f_1(x,y,z)=x{1-y+yz\over 1-z+zx}.
$$
The other functions $f_2,\ f_3$ are determined from (\ref{r,s}) and we obtain the 3 dimensional Lotka-Volterra (3dLV) map
\begin{equation}
X=x{1-y+yz\over 1-z+zx},\quad Y=y{1-z+zx\over 1-x+xy},\quad Z=z{1-x+xy\over 1-y+yz}.
\label{LV}
\end{equation}
If we had chosen another function for $f_1$ we would have a different map but the same invariants.

After the elimination of $y$ and $z$ from (\ref{LV}) by using (\ref{r,s}) we obtain a biquadratic map of (\ref{S(X,x)=0}) with
$$
a=r+1,\quad b=s-2r-1,\quad c=r-s,
$$
\begin{equation}
d=s^2+rs+5r-2s+1,\quad e=-r(s+1),\quad f=0.
\label{LV parameters}
\end{equation}
In these formulae $(r,s)=(H_1,H_2)$ are the values of the constants of the map (\ref{LV}).

We can carry out similar calculations for the level set
$$
\varphi_1=x_1x_2x_3x_4-H_1,\quad \varphi_2=(1-x_1)(1-x_2)(1-x_3)(1-x_4)-H_2,
$$
\begin{equation}
\varphi_3=(1-x_2x_4)(1-x_1x_3)-H_3,
\label{invariants of Painleve}
\end{equation}
and the function
$$
f_1(\bm{x})=x_1{1-x_2+x_2x_3-x_2x_3x_4\over 1-x_4+x_4x_1-x_4x_1x_2}.
$$
From (\ref{invariants of Painleve}) we obtain the map
\begin{eqnarray}
X_1:=x_1{1-x_2+x_2x_3-x_2x_3x_4\over 1-x_4+x_4x_1-x_4x_1x_2},&\quad &
X_2:=x_2{1-x_3+x_3x_4-x_3x_4x_1\over 1-x_1+x_1x_2-x_1x_2x_3},\nonumber\\
X_3:=x_3{1-x_4+x_4x_1-x_4x_1x_2\over 1-x_2+x_2x_3-x_2x_3x_4},&\quad &
X_4:=x_4{1-x_1+x_1x_2-x_1x_2x_3\over 1-x_3+x_3x_4-x_3x_4x_1},
\label{Painleve V}
\end{eqnarray}
which is known as the discrete time Painlev\'e V. The reduction of this map yields again a biquadratic map (\ref{S(X,x)=0}) with the coefficients
\begin{eqnarray*}
a&=&(s+v-r+1)p+r-1,\\
b&=&(2r-s-v-2)p-2r-s-v+2,\\
c&=&(1-r)p+r+s+v-1,\\
d&=&4(1-r)p+2(r-1)(s+2)+(s+v)(4-s-v),\\
e&=&(2r+s+v-2)p+(s+1)(v-2r-1)+(v-3)(v-1),\\
f&=&-(r+rs+v-1)p+r+rs(r-v+1)-(v-1)^2,
\end{eqnarray*}
where $(r,s,v)=(H_1,H_2,H_3)$ and $p$ is a solution of
$$
p^2-(r-v+1)p+r=0.
$$

\section{Dimensional reduction of periodicity conditions}

We now study, in this section, general features  of periodicity conditions and present the proof of our theorem stated in \S 1.

We consider the periodicity conditions of period $n$,
\begin{equation}
X_j^{(n)}=x_j,\quad j=1,2,...,d.
\label{X^n=x}
\end{equation}
When the map has $p$ invariants, the number of independent conditions of (\ref{X^n=x}) is $d-p$. Let us denote by $\bm{\xi}=(\xi_1,\xi_2,...,\xi_{d-p})$ the variables which parameterize the variety $V(h)$ of (\ref{V(h)}) after the reduction of $p$ components of $\bm{x}$. The problem of finding periodic points is equivalent to finding the $p$th elimination ideal generated by some functions $\Gamma_n^{(\alpha)}$ satisfying
\begin{equation}
\Gamma_n^{(\alpha)}(h_1,h_2,...,h_p,\xi_1,\xi_2,...,\xi_{d-p})=0,\qquad \alpha=1,2,...,d-p.
\label{Gamma_n}
\end{equation}
This new set of equations (\ref{Gamma_n}) plays the role of the conditions (\ref{X^n=x}) after the dimensional reduction.

For arbitrary values of $h_1,h_2,...,h_p$, the equations (\ref{Gamma_n}) determine a set of isolated points, hence a variety of dimension 0, on $V(h)$ in general. We called in \cite{SSI} the periodicity conditions `uncorrelated' in this case. There are possibilities that the equations $\Gamma_n^{(\alpha)}=0$ impose relations on $h_1,h_2,...,h_p$ instead of fixing all $\xi_\alpha$'s. If $s$ is the number of independent constraints imposed on $h_i$'s, the same number of $\xi_\alpha$'s are left free. Accordingly the map of period $n$ is extended to an $s$ dimensional subvariety of $V(h)$. We call the periodicity conditions `correlated' in this case, and `fully correlated' when $s=d-p$.

We have, so far, assumed that the values of $h_1,h_2,...,h_p$ have been fixed and $\xi_1,...,\xi_{d-p}$ are determined by them. If the periodicity conditions are fully correlated, however, none of $\xi_{\alpha}$'s is constrained but only relations among the invariants $(h_1,h_2,...,h_p)$ are imposed. To emphasize the independence from the variables $\xi_1,\xi_2,...,\xi_{d-p}$, in this particular case, we denote by $\gamma_n^{(\alpha)}(h_1,h_2,...,h_p)$ the $d-p$ functions $\Gamma_n^{(\alpha)}(h_1,h_2,...,h_p,\xi_1,\xi_2,...,\xi_{d-p})$. Note that the periodicity conditions are relevant only to the relations among the invariants but not to their values. This means that, by substituting $H_i(\bm{x})$ for $h_i$ in $\gamma_n^{(\alpha)}(h)$, every point on the set defined by
$$
v(\langle\gamma_n\rangle)=\Big\{\bm{x}\Big|\ \gamma^{(\alpha)}(H_1(\bm{x}),H_2(\bm{x}),...,H_p(\bm{x}))=0,\ \ \alpha=1,2,...,d-p\Big\},
$$
is a periodic point of period $n$. We called $v(\langle \gamma_n\rangle)$ in \cite{SSI} an `invariant variety of periodic points'. We must emphasize that $v(\langle \gamma_n\rangle)$ is a subvariety of $\bm{C}^d$, but not of $V(h)$.

For the properties of $v(\langle \gamma_n\rangle)$ we simply summarize as follows.
\begin{enumerate}
\item
The dimension of $v(\langle \gamma_n\rangle)$ is $p$.
\item
Every point on $v(\langle \gamma_n\rangle)$ can be an initial point of the periodic map of period $n$.
\item
All images of the periodic map started from a point of $v(\langle \gamma_n\rangle)$ remain on it.
\item
$v(\langle \gamma_n\rangle)$ is determined by the invariants of the map alone.
\end{enumerate}
\vglue0.5cm

Having introduced some notions about periodicity conditions we are now going to establish the theorem stated in \S 1. To this end let us consider the periodicity conditions of period $k$ given by
\begin{eqnarray}
\gamma_k^{(\alpha)}(h_1,h_2,...,h_p)=0,&\quad& \alpha=1,2,...,s
\label{gamma_k}
\\
\Gamma_k^{(\alpha)}(h_{1},...,h_p,\xi_{s+1},...,\xi_{d-p})=0,&\quad& \alpha=s+1,...,d-p.
\label{Gamma_k}
\end{eqnarray}
They are fully correlated if $s=d-p$, correlated if $d-p\ge s>0$ and uncorrelated if $s=0$. The fully correlated periodicity conditions are possible only when $p\ge d/2$, because $p\ge s$ is true always.

We now assume that the periodicity conditions (\ref{Gamma_n}) of period $n\ne k$ are uncorrelated. Let $\xi_{s+1}(h),\xi_{s+2}(h),...,\xi_{d-p}(h)$ be solutions of (\ref{Gamma_k}) and substitute them to (\ref{Gamma_n})
\begin{equation}
\Gamma_n^{(\alpha)}(h_1,...,h_p,\xi_1,...,\xi_s,\xi_{s+1}(h),...,\xi_{d-p}(h))=0,\qquad \alpha=1,...,d-p.
\label{Gamma_n(h)}
\end{equation}
We consider (\ref{Gamma_n(h)}) as a set of $d-p$ equations to be solved for $\xi_1,...,\xi_s$ when the invariants are given. Since the number of the equations in (\ref{Gamma_n(h)}) is larger than the number of the unknown variables $\xi_1,...,\xi_s$, the existence of solutions to (\ref{Gamma_n(h)}) depends on the values of the invariants. From (\ref{gamma_k}) we can decide $s$ invariants, say $h_1,h_2,...,h_s$, in terms of the rest of them. If the number $p-s$ of the free invariants $h_{s+1},...,h_{p}$ exceeds $d-p-s$, thus $p\ge d/2$ holds, we obtain solutions of (\ref{Gamma_n(h)}) for $\xi_1,...,\xi_s$, because the algebraic equations (\ref{Gamma_n(h)}) always have solutions for the $d-p$ unknown variables $\xi_1,...,\xi_s,h_{s+1},...,h_{p}$. We now recall that the values of $\xi_1,\xi_2,...,\xi_{d-p}$ satisfying (\ref{Gamma_n(h)}) also satisfy (\ref{Gamma_k}), since $\xi_{s+1},...,\xi_{d-p}$ are common in both and $\xi_1,\xi_2,...,\xi_s$ can be chosen arbitrary in (\ref{Gamma_k}) if $s\ne 0$. In other words, if the periodicity conditions of period $n$ are uncorrelated, there exist points of period $n$ which coincide with points of period $k$ as long as $p\ge d/2$ and $s\ge 1$. This apparent contradiction to our assumption $n\ne k$ shows that a set of uncorrelated periodicity conditions is incompatible with $p\ge d/2$ when there exists a set of correlated periodicity conditions. If $p<d/2$, on the other hand, (\ref{Gamma_n(h)}) are not forced to share points of other periods, hence uncorrelated periodicity conditions are not excluded in general. From this result our theorem presented in \S 1 follows immediately.\\

We notice that our theorem does not exclude possibilities that all periodicity conditions are uncorrelated even though the map has invariants more than $d/2$. For example we discussed in \S 2 two dimensional maps (\ref{X=xy,Y=x+y-xy}) and (\ref{2d map}) which become the logistic map after the dimensional reduction using the invariants.

\section{Series of invariant variety of periodic points}

We have presented many examples of invariant variety of periodic points associated with some known integrable maps. They have been derived by solving the periodicity conditions one by one. The purpose of this section is to show that, for some simple cases, we can derive iteratively series of infinite number of $\gamma_n$'s.

First we study the two dimensional map (\ref{2d map}). After the elimination of $y$ there is only one periodicity condition $X^{(n)}=x$ for each period $n$. The variable $\xi$ that parameterizes $V(h)$ is $x$ in this case. The condition of period 2 is given by $\Gamma_2(h,x)=0$ with
$$
\Gamma_2(h,x)=h+1+chx{chx+bx-1-h\over 1-bx}
$$
When $c\ne 0$, there are two solutions corresponding to two isolated points. When $c=0$, {\it i.e.}, the case of the M\"obius map, the condition is fully correlated since
$$
\gamma_2(h)=h+1
$$
is a function of the invariant $h$ alone. By substituting the formula (\ref{H=y(1-bx)/(1-cx)}) for the invariant with $c=0$ we find the invariant variety of periodic points of period 2 as
$$
y(1-bx)+1=0.
$$

We can continue this to the period $n$ and obtain, in the case of $c=0$,
$$
\gamma_n(h)=\gamma_{n-1}(h)+h^{n-1}.
$$
Therefore we obtain
$$
\gamma_n(h)={h^{n}-1\over h-1},\qquad n=1,2,3,....
$$
The invariant variety is $(n-1)$fold as it is derived by the substitution of $y(1-bx)$ for $h$ in
\begin{equation}
h=e^{i2\pi k/n},\qquad k=1,2,...,n-1.
\label{h=-e^i 2pi k/n}
\end{equation}
\vglue0.5cm

The M\"obius map, which we have just discussed, is a unique rational one dimensional map which is integrable. If we consider higher dimensions we have shown in \S 2 that there are many rational maps which are integrable and reduce to the biquadratic map (\ref{S(X,x)=0}) of one dimension. In the rest of this section we study this map.

The problem of finding the image $Q(x)$ of the second iteration of the map (\ref{S(X,x)=0}) is equivalent to finding an elimination ideal generated by $S(Q,X;\bm{q})$ and $S(X,x;\bm{q})$. After some manipulation we obtain an equation $W_2(Q,x)=0$, where
\begin{equation}
W_2(Q,x)=(Q-x)^2S(Q,x;\bm{q}_2),
\label{(Q-x)^2S_2(Q,x)=0}
\end{equation}
which fixes $Q$ as a function of $x$. The new parameters $\bm{q}_2=(a_2,b_2,c_2,d_2,e_2,f_2)$ are given by
\begin{eqnarray}
a_2&:=& (ae-cb)^2-(ad-2ac-b^2)(be-cd+2c^2),\nonumber\\
b_2&:=& (ae-cb)(2af-be+cd-4c^2)-(ad-2ac-b^2)(bf-ce),\nonumber\\
c_2&:=&(af-c^2)^2-(ae-bc)(bf-ce),\label{a_2,...,f_2}\\
d_2&:=&4(af-c^2)^2-2(ae-bc)(bf-ce)-(be-cd+2c^2)^2\nonumber\\
&&
\quad-(ad-2ac-b^2)(df-2cf-e^2),\nonumber\\
e_2&:=& (fb-ce)(2af-be+cd-4c^2)-(fd-2fc-e^2)(ea-cb),\nonumber\\
f_2&:=& (fb-ce)^2-(fd-2fc-e^2)(be-cd+2c^2).\nonumber
\end{eqnarray}

It is not difficult to understand the meaning of the expression of the right hand side of (\ref{(Q-x)^2S_2(Q,x)=0}). The biquadratic map is symmetric under the exchange of the arguments, as it is manifest in (\ref{S(X,x)}). Due to this symmetry, the backward map $X=X^{(-1)}(x)$ solves $S(X,x;\bm{q})=0$ when the foreward map $X^{(1)}(x)$ solves it. The substitution of $X=X^{(1)}(x)$ to $S(Q,X;\bm{q})$ for $X$ yields $Q=X^{(2)}(x)$ and $Q=x$ as its solutions, while the substitution of $X=X^{(-1)}(x)$ yields $Q=x$ and $Q=X^{(-2)}(x)$. Therefore the function $W_2(Q,x)$ must be proportional to $(Q-x)^2$. Moreover the elimination of $X$ from $S(Q,X;\bm{q})$ and $S(X,x;\bm{q})$ makes $W_2(Q,x)$ quartic both in $Q$ and $x$. Thence another factor in $W_2(Q,x)$ must be symmetric and also quadratic in $Q$ and $x$. The most general function satisfying these conditions is the biquadratic function $S(Q,x;\bm{q}_2)$ itself. The parameters $\bm{q}_2$ can be fixed by comparing both sides of (\ref{(Q-x)^2S_2(Q,x)=0}) and are given by (\ref{a_2,...,f_2}). Note that $S(Q,x;\bm{q}_2)=0$ is solved by $X^{(\pm 2)}(x)$.\\

Now we suppose that the $k$th image $X^{(k)}$ of the map (\ref{S(X,x)=0}) is found up to $k=n$ by solving $S(Q,x;\bm{q}_k)=0,\ k=2,3,...,n$ for $Q$. 
The $(n+1)$th image $X^{(n+1)}$ of $x$ must be obtained by solving an equation $W_{n+1}(Q,x)=0$ which is derived from $S(Q,X;\bm{q}_n)=0$ and $S(X,x;\bm{q})=0$ after the elimination of $X$. Now by the substitution of $X^{(\pm 1)}(x)$ to $S(Q,X;\bm{q}_n)=0$ for $X$ we obtain solutions $X^{(n\pm 1)}(x)$ and $X^{-(n\mp 1)}(x)$. We know already $X^{\pm(n-1)}(x)$ as the solutions of $S(Q,x;\bm{q}_{n-1})=0$. Therefore $W_{n+1}(Q,x)$ must have $S(Q,x;\bm{q}_{n-1})$ as a factor. From the same argument in the case of $n=2$ another factor must be $S(Q,x;\bm{q}_{n+1})$, hence we have
\begin{equation}
W_{n+1}(Q,x)=S(Q,x;\bm{q}_{n-1})S(Q,x;\bm{q}_{n+1}).
\label{W_n+1}
\end{equation}
The new coefficients $\bm{q}_{n+1}$ will be obtained by comparing both sides of (\ref{W_n+1}) as functions of $\bm{q}$ and $\bm{q}_{n}$ iteratively. Using the notation $(g_\wedge g')_n=gg'_n-g'g_n$, the results are as follows:
\begin{eqnarray}
a_{n+1}&=&{1\over a_{n-1}}\Big((a_\wedge c)_n^2-(a_\wedge b)_n(b_\wedge c)_n\Big),\nonumber\\
b_{n+1}&=&{1\over a_{n-1}}\Bigg({b_{n-1}\over a_{n-1}}\Big((a_\wedge b)_n(b_\wedge c)_n-(a_\wedge c)_n^2\Big)
+(a_\wedge c)_n\Big((a_\wedge e)_n+2(b_\wedge c)_n\Big)\nonumber\\
&&\qquad-\ {1\over 2}\Big((a_\wedge b)_n(b_\wedge e)_n-(a_\wedge b)_n(c_\wedge d)_n+(a_\wedge d)_n(b_\wedge c)_n\Big)\Bigg),\nonumber\\
c_{n+1}&=&{1\over 2c_{n-1}}\Bigg((ce_n-bf_n)(ae_n-bc_n)+(cb_n-ea_n)(fb_n-ec_n)\nonumber\\
&&
\qquad\qquad+(af_n-cc_n)^2+(fa_n-cc_n)^2\Bigg),\label{n+1th parameters}\\
d_{n+1} &=& {1\over d_{n-1}}\Bigg(-f_{n-1}a_{n+1}-a_{n-1}f_{n+1}-4b_{n-1}e_{n+1}-4e_{n-1}b_{n+1}+(a_\wedge f)_n^2+(c_\wedge d)_n^2\nonumber\\
&&
\qquad-(a_\wedge b)_n(e_\wedge f )_n-(b_\wedge c)_n(c_\wedge e)_n+(a_\wedge d)_n(d_\wedge f)_n+2(b_\wedge e)_n(a_\wedge f)_n\nonumber\\
&&
\qquad-\Big(3(c_\wedge e)_n-(b_\wedge f)_n-(d_\wedge e)_n\Big)\Big(3(b_\wedge c)_n-(a_\wedge e)_n-(b_\wedge d)_n\Big)\nonumber\\
&&+2\Big((a_\wedge d)_n-(a_\wedge c)_n\Big)\Big((c_\wedge f)_n-(d_\wedge f)_n\Big)+2\Big((b_\wedge c)_n+(a_\wedge e)_n\Big)\Big((b_\wedge f)_n+(c_\wedge e)_n\Big)\Bigg),
\nonumber\\
e_{n+1}&=&{1\over f_{n-1}}\Bigg({e_{n-1}\over f_{n-1}}\Big((f_\wedge e)_n(e_\wedge c)_n-(f_\wedge c)_n^2\Big)
+(f_\wedge c)_n\Big((f_\wedge b)_n+2(e_\wedge c)_n\Big)\nonumber\\
&&\qquad-\ {1\over 2}\Big((f_\wedge e)_n(e_\wedge b)_n-(f_\wedge e)_n(c_\wedge d)_n+(f_\wedge d)_n(e_\wedge c)_n\Big)\Bigg),\nonumber\\
f_{n+1}&=&
{1\over f_{n-1}}\Big((f_\wedge c)_n^2-(f_\wedge e)_n(e_\wedge c)_n\Big).\nonumber
\end{eqnarray}

\vglue0.5cm

Our problem of studying the behaviour of the map (\ref{S(X,x)}) has been converted to studying the behaviour of the parameters $\bm{q}_n=(a_n,b_n,...,f_n)$. These parameters determine a trajectory of iteration of the map (\ref{S(X,x)=0}):
$$
\cdots \longleftarrow X^{(-2)}\longleftarrow X^{(-1)}\longleftarrow x\longrightarrow X^{(1)}\longrightarrow X^{(2)}\longrightarrow \cdots .
$$
Namely we can consider (\ref{n+1th parameters}) as a map $\bm{q}_n\ \rightarrow\ \bm{q}_{n+1}$ in $\bm{C}^6$.\\

The periodicity conditions of period $n$ for the map $x\rightarrow X$ defined by (\ref{S(X,x)=0}) are satisfied if the parameters satisfy
\begin{equation}
\bm{q}_n(\bm{q})=\bm{q}.
\label{p_n=p_1}
\end{equation}
Despite the complicated expression of the relation (\ref{n+1th parameters}), we observe a special dependence on the $n$th parameters $\bm{q}_n$. Besides $c_{n+1}$, the dependence of the $(n+1)$th parameters on the $n$th ones is always in the form $(g_\wedge g')_n=gg'_n-g'g_n$. They all vanish simultaneously when the periodicity conditions (\ref{p_n=p_1}) are `fully correlated'. In other words if there exists a function $\gamma_{n+1}(\bm{q})$ such that
\begin{eqnarray}
\bm{q}_n(\bm{q})&=&\bm{q}+\gamma_{n+1}(\bm{q})\hat{\bm{q}}_n(\bm{q})
\label{a_n=a+gamma a}
\end{eqnarray}
so that (\ref{p_n=p_1}) is satisfied by a single condition $\gamma_{n+1}(\bm{q})=0$. Here we use the same notation $\gamma_n$, which we used already as a function of $h$ in higher dimension. Although they arise in different contexts they turn out to be the same object as we will see later.

When (\ref{a_n=a+gamma a}) holds, the equation $S(Q,x;\bm{q}_{n+1})=0$ can be written as
\begin{equation}
c_{n+1}(Q-x)^2+\gamma_{n+1}^2(\bm{q})K_{n+1}(Q,x)=0.
\label{c_n(Q-x)^2+gamma_nK_n(Q,x)=0}
\end{equation}
Here 
\begin{eqnarray}
K_{n+1}(Q,x)&=&\hat a_{n+1}Q^2x^2+\hat b_{n+1}(Q+x)Qx+\hat d_{n+1}Qx+\hat e_{n+1}(Q+x)+\hat f_{n+1},
\label{K_n}
\end{eqnarray}
and $\hat a_{n+1}$, for instance, is obtained from $a_{n+1}$ simply replacing $(g_\wedge g')_n$ by $(\hat g_\wedge{\hat g}')_n$. If $Q$ is a point of period $n+1$, the first term of (\ref{c_n(Q-x)^2+gamma_nK_n(Q,x)=0}) vanishes. Hence the periodicity condition requires for the second term to vanish. This is certainly satisfied for arbitrary $x$ if $\gamma_{n+1}(\bm{q})=0$, namely when the periodicity conditions for the parameters $\bm{q}_n$ are fully correlated. The other possible solutions obtained by solving $K_{n+1}(x,x)=0$ will not correspond to the points of period $n+1$, but represent the fixed points or the points with periods which divide $n+1$.\\

Let us present the functions $\gamma_n(\bm{q})$ explicitly in the cases of small number of $n$. We can show
\begin{eqnarray*}
(a_\wedge b)_2&=&(af-eb-3c^2+cd)(2a^2e-abd+b^3)\\
(a_\wedge c)_2&=&(af-eb-3c^2+cd)(a^2f+ac^2-acd+b^2c)\\
(b_\wedge c)_2&=&(af-eb-3c^2+cd)(2ace-abf-bc^2)\\
&\vdots&\\
(e_\wedge f)_2&=&(af-eb-3c^2+cd)(edf-e^3-2bf^2),
\end{eqnarray*}
from which we find
\begin{equation}
\gamma_3(\bm{q})=af-be-3c^2+cd.
\label{gamma_3(p)}
\end{equation}
If we further calculate the cases of $n=3,4$,
\begin{eqnarray}
\gamma_4(\bm{q})&=&2acf-adf+b^2f+ae^2-2c^3+c^2d-2bce,
\nonumber\\
\gamma_5(\bm{q})&=&
a^3f^3+\Big(-cf^2d+2cfe^2+fde^2-3ebf^2-e^4-c^2f^2\Big)a^2\nonumber\\
&&+\Big(-13c^4f+18c^3fd+de^3b+2cf^2b^2+7dc^2e^2-ce^2d^2-2ce^3b\label{gamma_4(p)}\\
&&+2c^2feb-7fd^2c^2-14c^3e^2+cd^3f+fb^2e^2+f^2db^2-ebd^2f\Big)a\nonumber\\
&&
-cd^2b^2f-b^3e^3-4c^3deb+cdb^2e^2+13ec^4b-f^2b^4+7fb^2c^2d\nonumber\\
&&
+c^4d^2-5c^5d+5c^6-2fb^3ec-e^2c^2b^2+eb^3df-14fb^2c^3.\nonumber
\end{eqnarray}
The formula (\ref{n+1th parameters}) enables us to continue finding a series of $\gamma_n(\bm{q})$ systematically. From the way of this construction it is apparent that all $\gamma$'s are functions of the invariants alone if one of them is so, a fact being consistent with our theorem.\\

The 3-dim Lotka-Volterra map after reduction to one dimension is a special example of the biquadratic map (\ref{S(X,x)}). The parameters have been specified by (\ref{LV parameters}). From the general argument we obtain the result of first iteration simply substituting these data into (\ref{a_2,...,f_2}). They are given by
\begin{eqnarray}
a_2&=&(s+1)^2s(r^2-rs^2-s-3rs),\nonumber\\
b_2&=&(s+1)^2s(2r^2s+s+5rs-2r^2-r^3-s^2),\nonumber\\
c_2&=&(s-r)s(r+1)(s^2-r^2s-3rs-r)\label{LV p_2},\\
d_2&=&(s+1)^2s(2rs^2+2s^2-3r^2s-8rs-s+r^3s+r^4+5r^3+6r^2-s^3),\nonumber\\
e_2&=&(s+1)^2rs(s^2-rs+s-2r-r^3-2r^2),\nonumber\\
f_2&=&(s+1)^2r^2s(r^2-rs+r+s^2+s+1).\nonumber
\end{eqnarray}
Notice that all parameters apart from $c_2$ are proportional to a common factor $(s+1)^2$. This is a result which we cannot derive generally from (\ref{a_2,...,f_2}) since they are not factorized. On the other hand this factor is exactly the one we expect from our previous result \cite{SSI}, although this is a special behaviour of 3dLV map.

The periodicity conditions of period 3 and higher can be derived similarly and the associated functions $\gamma_3$, $\gamma_4$ and $\gamma_5(\bm{q})$ can be read off from (\ref{gamma_3(p)}) and (\ref{gamma_4(p)}) directly. We find
\begin{eqnarray*}
\gamma_3(r,s)&=&r^2+s^2-rs+r+s+1,\\
\gamma_4(r,s)&=&3rs+s+s^3-3s^2r+r^3s+6r^2s-r^3,\\
\gamma_5(r,s)&=&r^3s^4-r^3s^2-6r^4s^5+10r^3s^6+3s^5r+s^6+s^5+3r^4s^4-3r^5s^3-6r^4s^3\\
&&
-r^6s^3+3r^5s^4+s^4+21s^4r^2+6s^4r+r^3s^7+s^7+27s^5r^2-3s^6r-r^3s^5\\
&&
+21r^2s^6-10r^3s^3-6rs^7+s^8.
\end{eqnarray*}
These include again precisely the conditions expected from our direct calculations in \cite{SSI}. 

\section{Transition to integrable regime}

The method of generating maps, developed in \S 2, enables us to study maps having invariants from a general point of view. For example this scheme provides us a map which becomes either integrable or nonintegrable upon reduction to one dimension, but shares invariants in common. Such a map will be convenient for the study of the transition between integrable and nonintegrable regimes. 

The map presented in (\ref{2d map}) is one such example. By adjusting properly the parameters the map can be shifted from nonintegrable to integrable. This is seen clearly from the expression of one dimensional map (\ref{X=hx(1-cx)/(1-bx)}), but not obvious from the expression of (\ref{2d map}) written as a two dimensional map. We are going to study this particular map in this section in detail, since it is simple but rich enough. The purpose of this section is to explore how this transition takes place as $c$ approaches zero continuously. We would like to know where the periodic points go in this limit. For this purpose it is sufficient to study the map (\ref{X=hx(1-cx)/(1-bx)}). 

Since one dimensional rational maps of degree 2 have been studied in the literature it will be convenient to discuss a map of the normal form\cite{FM},
\begin{equation}
Z=z{h' +z\over 1+hz},
\label{Z}
\end{equation}
after conjugating by the transformation $x\rightarrow z$
\begin{equation}
z={1-h'\over 1-h}+{1-h\over h(b-c)}\ {1\over x}.
\label{z--x}
\end{equation}
Here the parameters are related by
$$
hh'=1+{c\over b-c}(1-h)^2,
$$
hence the integrable limit $c=0$ corresponds to $hh'=1$.

Some basic data of the map (\ref{Z}) are as follows. The fixed points are at $0,\infty , z_p={1-h'\over 1-h}$ and their multipliers are given, respectively, by
$$
h',\quad h,\quad {2-h-h'\over 1-hh'}.
$$
This map has two critical points at 
$$
z_\pm=-{1\over h}\pm {\sqrt{1-hh'}\over h},
$$
where the multipliers vanish. \\

Repeating the map $n$ times the image of $z$ can be written as
\begin{eqnarray}
Z^{(n)}&=&{1\over h}Z^{(n-1)}{h'+Z^{(n-1)}\over h^{-1}+Z^{(n-1)}}
\label{Z^(n)1}\\
&=&
{1\over h^n}z\left({h'+z\over h^{-1}+z}\right)\left({h'+Z^{(1)}\over h^{-1}+Z^{(1)}}\right)\left({h'+Z^{(2)}\over h^{-1}+Z^{(2)}}\right)\cdots\left({h'+Z^{(n-1)}\over h^{-1}+Z^{(n-1)}}\right).
\label{Z^(n)2}
\end{eqnarray}
If $k_{n-1}$ is the degree of the rational polynomial $Z^{(n-1)}$, we see from (\ref{Z^(n)1}) that $k_n=2k_{n-1}$, hence $k_n=2^n$. Periodic points are obtained by solving $Z^{(n)}=z$. When $n$ is a prime number, the number $\#_n$ of the periodic points of period $n$ is $k_n-2=2^n-2$. Here subtraction of 2 corresponds to two fixed points at 0 and $z_p$. If $n$ is not prime but has divisors $\nu_1,\nu_2,\cdots,\nu_r$ they must be also subtracted. We thus find the number of periodic points of period $n$ as
\begin{eqnarray}
\#_n&=&2^n-\#_{\nu_1}-\#_{\nu_2}-\cdots -\#_{\nu_r}-2
\nonumber\\
&=&
2^n-2^{\nu_1}-2^{\nu_2}-\cdots -2^{\nu_r}+2(r-1),
\label{number_n}
\end{eqnarray}
which increases as fast as 
$$
(\#_2,\#_3,\#_4,\cdots)=(2,6,12,30,48,126,240,504,\cdots).
$$

In the integrable limit $hh'=1$, this increase ceases owing to the cancellation of factors in the numerator and the denominator as seen in the second expression (\ref{Z^(n)2}), and the map becomes
\begin{equation}
Z^{(n)}=h^{-n}z.
\label{Z=h^-nz}
\end{equation}
It is apparent that there is no periodic point unless $h$ satisfies (\ref{h=-e^i 2pi k/n}). All isolated points suddenly disappear. In order to uncover this trick of the transition we must know where they were right before they disappeared. After some manipulation we find that the right hand side of (\ref{Z^(n)2}) admits the following expression 
$$
Z^{(n)}=
{z\over h^n}\ {\displaystyle{\prod_{k=-1}^{n-2}(z+h^k)^{2^{n-k-2}}+(hh'-1)P_n}\over
\displaystyle{\prod_{k=-1}^{n-2}(z+h^k)^{2^{n-k-2}}+(hh'-1)Q_n}}
$$
where $P_n$ and $Q_n$ are polynomials of $z$. The periodicity condition $Z^{(n)}=z$ requires
\begin{equation}
(h^n-1)\prod_{k=-1}^{n-2}(z+h^k)^{2^{n-k-2}}
=
(hh'-1)(P_n-h^nQ_n)
\label{periodicity cond for small}
\end{equation}
to hold. When $hh'=1$ we obtain solutions at
\begin{equation}
z=-1/h,-1,-h,\cdots,-h^{n-2}
\label{-1/h,-1,-h..}
\end{equation}
if $h^n\ne 1$. 

These points are, however, certainly not periodic solution of (\ref{Z=h^-nz}). They are ``fossils''. Suppose $hh'-1$ is very small but not zero. Then the periodicity condition (\ref{periodicity cond for small}) is satisfied iff $z$ is very close to one of (\ref{-1/h,-1,-h..}). Therefore the periodic points were in the neighbourhood of these points right before $hh'$ reached 1. \\


The behaviour of the iteration of a map is strongly controlled by the existence of repulsive periodic points. The Julia set is the closure of the set of repulsive periodic points of a map, and is a source of chaotic orbits of the iteration. After the transition to the integrable map it is expected to disappear or become neutral points. Therefore it is quite useful to see how the Julia set behaves when a map is changed from nonintegrable to integrable.

There are several equivalent ways of defining the Julia set. We adopt the following one which is convenient for our purpose\cite{Devaney}.

\noindent
Definition:

{\it 
If $f$ is a map with inverse map $f^{-1}$ and $z_0$ is a repulsive periodic point, then the Julia set $J(f)$ of $f$ is defined by:
\begin{equation}
J(f)=\left\{z\left|\bigcup_{l=0}^\infty\right. f^{-l}(z_0)\right\}.
\end{equation}
}

This tells us that starting from one of the repulsive points we will find all points in the Julia set simply by repeating the inverse map. Since this definition, applicable to one dimensional maps, requires the data of the inverse map, we expect to supply information complementary to our results in the previous subsection. In fact we shall prove that all repulsive points of the Julia set converge uniformly to the countable set of neutral points (\ref{-1/h,-1,-h..}) in the integrable limit\cite{YS,SSS}.

The inverse map of (\ref{Z}) is
\begin{equation}
f^{-1}(z)={1\over 2}\left(hz-h'\pm\sqrt{(hz+h')^2+4(1-hh')z}\right).
\label{f^-1}
\end{equation}
Starting from a repulsive fixed point, say 0 when $h'>1$, the $n$th inverse map adds $2^n$ points to $J(f)$, and the total number of points in the Julia set increases as fast as $2^{n+1}-1$.

The situation, however, changes significantly when $hh'=1$. In this particular case the inverse map becomes
\begin{equation}
f^{-1}(z)=\left\{\matrix{hz,\cr -\ 1/h.\cr}\right.
\label{hz, -1/h}
\end{equation}
Note that, when $hh'=1$, $- 1/h$ is the critical point of the map (\ref{Z}) and becomes neutral.
Repeating the map $n$ times starting from 0 the maximum number of different points is $n+1$, consisting of
\begin{equation}
J_n:=\Big(0,-h^{-1},-1,-h,-h^2,\cdots, -h^{n-2}\Big).
\label{J_n}
\end{equation}
We have thus reproduced exactly the same series (\ref{-1/h,-1,-h..}) we found before. How they could be periodic points? We can answer this question now, which we could not see from the observation of the forward map alone. They are periodic in the sense that at every step of the map (\ref{hz, -1/h}) there is a possibility to return to the value $-\ 1/h$ and then continue the map $-1,-h,\cdots$ again. The Julia set in the limit of $hh'\rightarrow 1$ is given by $J_\infty$ and all periodic points of the map (\ref{Z}) must approach it.\\

Now we may ask how they approach $J_\infty$. This problem has been discussed in \cite{YS} some time ago. There the following theorem was proved.

{\it All points of the Julia set of the map (\ref{Z}) approach $J_\infty$ uniformly in the integrable limit. }

The proof is quoted briefly in Appendix from \cite{YS}.

\section{Concluding remarks}

By studying periodicity conditions of a map on $\bm{C}^d$ with $p$ invariants we proved our theorem presented in \S 1. We can rephrase it as
\begin{itemize}
\item
{\it When $p<d/2$, there is no invariant variety of periodic points.}
\item
{\it When $p\ge d/2$, a set of isolated periodic points and a variety of finite dimension do not exist in one map simultaneously.}
\end{itemize}

Because a precise notion of integrability of a map has not been known, the above statements do not provide criteria to distinguish an integrable map and a nonintegrable one. There are, however, some reasons to believe the following correspondence to hold:\\

\begin{tabular}{llcl}
1)&series of invariant variety of periodic points &$\Leftrightarrow$& integrable map\\
2)&fractal sets of isolated periodic points
&$\Leftrightarrow$& nonintegrable map\\
\end{tabular}

\vglue0.5cm
If we accept this correspondence the above statements become criteria for the integrability of a map.

We have shown, in our previous paper\cite{SSI}, many examples of invariant variety of periodic points which are derived from known integrable maps. On the other hand a fractal set of isolated periodic points appears typically in a nonintegrable map and becomes a source of a chaotic trajectory of iteration of the map. The study of transition of a logistic map to the M\"obius map, which we discussed in \S 5, shows an example of the above correspondence. It will become clear if we summarize the results of the two dimensional map (\ref{2d map}) as follows:\\

\begin{tabular}{|c|c|c|c|}
\hline
&periodicity conditions&periodic points& 1-dim reduction\\
\hline
$c\ne 0$&uncorrelated &isolated& logistic map\\\hline
$c=0$&fully correlated& invariant variety& M\"obius map\\
&$\gamma(h)=1+h+\cdots+h^{n-1}$& $\{(x,y)|y(1-bx)=e^{i2\pi k/n}\}$&\\
\hline
\end{tabular}
\vglue1cm

In the case of Hamilton dynamics with $N$ freedom, the existence of $N$ involutive invariants including the Hamiltonian guarantees integrability of the system. It will be useful to see if there is any correspondence of our theorem with that of Liouville-Arnold. If we count the momenta in the Hamilton system as independent variables, we may associate $2N$ with the dimension $d$ of the map. The number of the invariants, which are necessary and sufficient for the Hamilton system to be integrable, is $d/2$. This is exactly the critical number of the invariants $p$ of the map, which prohibits the coexistence of sets of uncorrelated and  correlated periodicity conditions. 

There is, however, an important difference between a continuous time evolution and a discrete map. A set of uncorrelated periodicity conditions, hence a series of isolated periodic points, is possible in a map even if there are invariants more than $N=d/2$. The map (\ref{2d map}) discussed above is an example. In general $d-1$ invariants are not sufficient for a map on $\bm{C}^d$ to be integrable after the reduction to one dimension. This should be contrasted with the fact that every first order ordinary differential equation is integrable. It is well-known that a proper continuous limit of the logistic map is integrable.\\

\noindent
{\bf Acknowledgement}

We would like to express our thanks to Professor Martin Guest who gave us many important suggestions to improve this paper.


\noindent
{\bf Appendix}

Here we quote briefly the proof of the Theorem from \cite{YS}. We can write (\ref{f^-1}) as
\begin{equation}
f^{-1}(z)=\left\{ \begin{array}{lcl}
A(z)&:=&hz+E(z)\cr
B(z)&:=&-h'-E(z)\cr
\end{array}\right.
\label{A,B}
\end{equation}
$$
E(z):={1\over 2}(hz+h')\left(\sqrt{1-{4z\epsilon\over (hz+h')^2}}-1\right),
$$
where we defined $\epsilon=hh'-1$. $E(z)$ vanishes for small values of $\epsilon$. In fact we can show
$$
|E(z)|\le R_\epsilon,
$$
for all values of $\epsilon$. Here $R_\epsilon$ is given by
$$
R_\epsilon={\sqrt 2+1\over |h|}\sqrt{|\epsilon|}\Big(\sqrt{|\epsilon|}+\sqrt{|\epsilon+1|}\Big),
$$
which tends to zero like $\sqrt{|\epsilon|}$ as $\epsilon$ approaches to zero. The proof of our claim follows from the facts:
\begin{enumerate}
\item
Using $A$ and $B$ defined by (\ref{A,B}) the $n$th iteration of $f^{-1}$ yields
\begin{equation}
f^{-n}(z)=\Big\{ A^{\nu_1}B^{\nu_2}A^{\nu_3}\cdots B^{\nu_n}(z)
\Big| 
\nu_1+\nu_2+\cdots+\nu_n=n\Big\}.
\label{f^-n(z)}
\end{equation}
\item
For any $W$ an elememt of the form $A^sBW,\ s=0,1,2,\cdots, n-1$ in (\ref{f^-n(z)}) lies in the neighbourhood of $-h^sh'$:
\begin{equation}
|A^s(BW)+h^sh'|=\Bigg|-h^sE(W)+\sum_{k=0}^{s-1}h^kE(A^{s-k-1}BW)\Bigg|<{1-|h|^s\over 1-|h|}R_\epsilon,
\label{|A^s(BW)+h^sh'|}
\end{equation}
$$
s=0,1,2,...,n-1.
$$
Since we assumed $|h'|>1,\ |h|<1$ is satisfied as long as $\epsilon$ is small. Hence the right hand side of (\ref{|A^s(BW)+h^sh'|}) is finite for all $s$.
\item
From 1. and 2. we conclude
$$
\Big|f^{-n}(z_0)-(-h^nh')\Big|<{1\over 1-|h|}R_\epsilon\quad\longrightarrow\quad 0,\qquad ^{\forall}n,\quad \epsilon\rightarrow 0.
$$
In our present problem the Julia set is the collection of all $f^{-n}(0)$, hence the claim is justified.
\end{enumerate}

\end{document}